\documentclass[epj]{svjour}

\usepackage{graphics}

\begin{document}

\title{Accumulation of chromium metastable atoms into an Optical Trap}

\author{R. Chicireanu, Q. Beaufils, A. Pouderous, B. Laburthe-Tolra, E. Mar\'echal, L. Vernac, J.-C. Keller, and O. Gorceix}
\institute{Laboratoire de Physique des Lasers, CNRS UMR 7538,
Universit\'e Paris 13, 99 Avenue J.-B. Cl\'ement, 93430
Villetaneuse, France}
\date{}

\abstract{ We report the fast accumulation of a large number of
metastable $^{52}$Cr atoms in a mixed trap, formed by the
superposition of a strongly confining optical trap and a quadrupolar
magnetic trap. The steady state is reached after about 400 ms,
providing a cloud of more than one million metastable atoms at a
temperature of about 100 $\mu K$, with a peak density of $10^{18}$
atoms.m$^{-3}$. We have optimized the loading procedure, and
measured the light shift of the $^5$D$_4$ state by analyzing how the
trapped atoms respond to a parametric excitation. We compare this
result to a theoretical evaluation based on the available
spectroscopic data for chromium atoms. 
}

\authorrunning{Chicireanu et al.}
\titlerunning{Accumulation of chromium metastable atoms into an Optical
Trap}
\maketitle

\section{Introduction}
\label{intro} Optical dipole traps have proved to be an important
tool to reach quantum degeneracy for a sample of neutral atoms since
the first realization of Bose Einstein Condensates (BEC) by all
optical means in 2001 \cite{BECCO2}. It took for example a crucial
part in the successful condensation of Cesium \cite{BECCs}, and
Chromium \cite{BECChrome}, for which high inelastic collision rates
have prevented to achieve BEC inside a magnetic trap
\cite{CsDalibard}. It was also necessary to use an optical trap (OT)
to reach BEC with Ytterbium atoms, which have no magnetic moment
(J=0) in the ground state \cite{YbTak}.

\begin{figure}
\resizebox{0.8\columnwidth}{!}{
\includegraphics{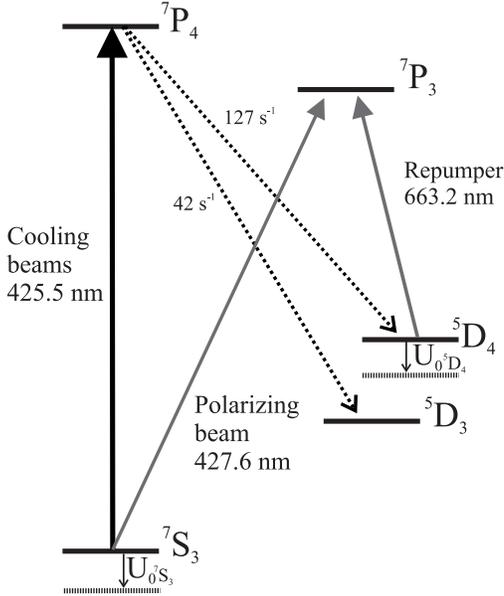}
} \caption{$^{52}$Cr atomic levels relevant for the study, and
optical lines used to cool, trap, repump or spin polarize the atoms.
From the excited $^7$P$_4$ state the atoms decay to metastable D
states at the indicated rates. The AC stark shifts induced by the
trapping IR laser as mentioned in the text are shown.}
\label{CrLevels}
\end{figure}

To reach BEC by evaporative cooling in a dipole trap, it is first
necessary to load a tightly confining OT with typically a few
million atoms. Given the available laser powers, such traps are very
small, and it is hard to load them from large volume Magneto Optical
Traps (MOT) or Magnetic Traps (MT). In the case of Cr, large
light-assisted inelastic losses \cite{CrMOTNIST,Chi1} make it
particularly difficult to load an OT directly from a MOT. However,
Cr atoms have a high magnetic moment, not only in the $^7$S$_3$
ground state (6 $\mu{_B}$, which makes Cr an interesting element to
explore dipolar effects in quantum degenerate gases
\cite{DegGasDip}), but as well in the metastable D states. This
latter property allows continuous accumulation of a large number of
cold metastable $^{52}$Cr atoms from a MOT into a MT
\cite{Chi1,Pfau1}: the atoms are optically pumped to the $^5$D$_4$
and $^5$D$_3$ metastable states by the cooling lasers (see Fig.
\ref{CrLevels}), and remain trapped in the quadrupolar MT formed by
the MOT coils. From this starting point, the successful strategy
followed to reach quantum degeneracy for Cr \cite{ProdCrBEC} has
been to accumulate the atoms in a Ioffe-Pritchard type elongated MT,
compress this trap and perform Doppler cooling in it after repumping
to the ground state, load a single axis OT from this MT after some
RF evaporative cooling, transfer the atoms into a crossed OT, and
finally evaporatively cool them in the crossed OT by decreasing the
laser power.

In this article we demonstrate that in 400 ms, more than one million
metastable Cr atoms can be directly accumulated in the OT created by
an intense laser beam crossing the MOT. After this accumulation, we
repump the atoms back to the $^7S_3$ ground state, switch the MOT
magnetic gradient off, and optically pump the atoms to the absolute
ground state $\left\vert^7S_3,M_J=-3\right\rangle$ with a laser beam
resonant with the 427.6 nm optical line (see Fig. \ref{CrLevels}).
Our results may thus provide a new strategy to produce quantum
degenerate gases of Cr atoms at a higher repetition rate, which may
also be relevant for other atoms with metastable states, such as Sr,
Er, or Yb.

Moreover, this article is to our knowledge the first to report the
optical trapping of neutral atoms in a metastable state. Even if the
optical trapping process is similar for metastable states and for
the ground state, the measurement of the OT depth in the D state
provides information on how this state couples to the other levels
of the complicated electronic structure of Cr, and can yield in
particular a determination of the polarizability of this state.

The paper is organized as follows. We first describe the
experimental setup and characterize the OT. We then present our
experimental results, which are the study of the loading of the OT,
and the parametric excitation spectra of the OT loaded by atoms in
the metastable $^5$D$_4$ state. To take into account the trap
anharmonicity, we perform 3D Monte Carlo simulations described in
section 4. We thus deduce the light shifts for the $^5$D$_4$ and the
$^7$S$_3$ states. Finally, we calculate the light shifts of the
different Zeeman sublevels of the states of interest, using the
spectroscopic data available for the optical transitions of Cr, and
we conclude with a comparison with our measurements.

\section{Realization of the optical trap}

\subsection{Experimental setup}

The experiment starts with the production of a Cr atom beam from an
oven heated at $1500$ Celsius. The atoms are then decelerated as
they travel through a one-meter-long Zeeman Slower (ZS), and they
are trapped in a standard MOT. We typically obtain a Gaussian-shape
cloud with a $1/e$ radius of 100 $\mu m$, containing up to
$5\times10^{6}$ $^{52}$Cr atoms at a temperature of 100 $\mu K$
\cite{Chi1}.

In previous works, we accumulated metastable D states atoms in the
MT created by the MOT coils \cite{Chi1,Chi2}. Here we study the
loading of metastable atoms into the OT created by an horizontal
laser beam focused at the MOT and MT center. This laser is a
commercial 1075 nm 50 W Ytterbium fiber laser. After an optical
isolator and some optical beam shaping to focus the laser in an
Acousto Optic Modulator (AOM) and re-collimate it, we have up to 35
W available for atom trapping. We focus the laser at the MOT
position with a plano convex lens of focal length $f_{IR}$. A weak
part of the IR laser transmitted through a back-side-polished
dielectric mirror is sent to a lens identical to the one producing
the OT, so that we can measure parameters very close to the ones of
the laser beam focused in the vacuum chamber. We obtained a waist of
$55(41)\pm5$ $\mu m$, and a Rayleigh length of $5.5(3)\pm.5$ mm with
$f_{IR}$=20 (respectively 15) cm.

In order to characterize the OT, we capture images of the atomic
cloud using a standard absorption technique. The 100 $\mu s$ imaging
beam pulse is resonant with the $^7$S$_3\rightarrow^7$P$_4$
transition ($\lambda$=425.5 nm, $I_{sat}=8$ mW.cm$^{-2}$), has an
intensity of 0.08 $I_{sat}$, and is circularly polarized. Its
propagation axis lies in the horizontal plane, and is almost
orthogonal to the ZS axis (see Fig. \ref{OTandImaging}). It makes an
angle $\alpha=7{{}^\circ}$ with the IR beam, which allows us to
monitor the expansion of the cloud along the weak trapping $z'$ axis
of the OT.

Before taking an image, the atoms are repumped to the $^7$S$_3$
ground state, via the $^5$D$_4\rightarrow^7$P$_3$ transition at
663.2 nm. For that purpose, we use a commercial monomode extended
cavity diode laser. A 2 mW-power and 5 ms long pulse is sufficient
to repump all the $^5$D$_4$ metastable atoms back in the ground
state. An other diode running at 654 nm can repump the atoms in the
$^5$D$_3$ state, so that we can monitor the accumulation of both
metastable states in the OT. However in this paper we will
concentrate on the $^5$D$_4$ state.

Then, a magnetic field of 2.8 Gauss is applied along $z$, and a 50
$\mu$s pulse at 427.6 nm, having an intensity of 0.5 $I_{sat}$ and
the same circular polarization than the imaging beam, polarizes the
atoms. Therefore during the imaging pulse the atoms cycle between
the fully stretched
$\left\vert^7S_3,M_J=-3\right\rangle\rightarrow\left\vert^7P_4,M_J=-4\right\rangle$
Zeeman sublevels and the average absorption cross section
$\sigma_{opt}$ can be set to $3\lambda^2/2\pi$ ($\lambda=425.5$ nm).

\subsection{Characterization of the dipole trap}

\begin{figure}
\resizebox{0.9\columnwidth}{!}{
\includegraphics{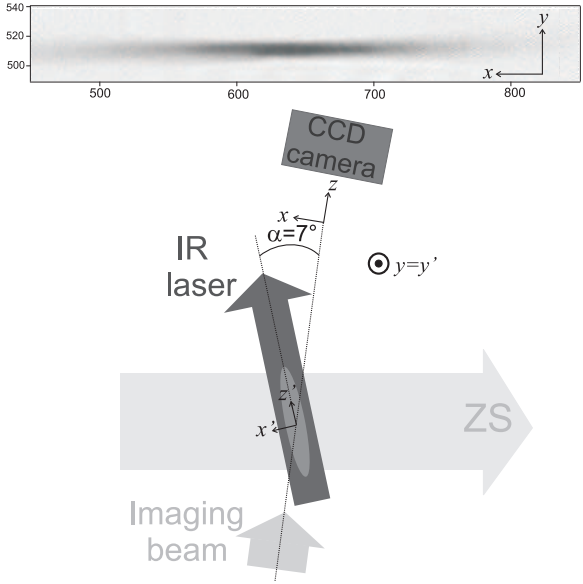}
} \caption{Bottom: schematic of the imaging system. ZS stands for
the Zeeman Slower beam. The imaging beam is partially absorbed by
the atoms in the OT created by the IR laser, and passes through a
1:1 imaging system (not shown) before reaching the CCD camera chip.
The axis of the two bases introduced in the text are shown. Top:
Optical Depth along $z$ after 400 ms of accumulation. The expansion
along the $x$ axis occurs during the switch-off time of the MOT
coils B field. The size of the pixels is 6.5 $\mu$m.}
\label{OTandImaging}
\end{figure}

The atomic cloud center is conjugated on the CCD chip plane by a 1:1
imaging system. We will deal with two bases: $(x,y,z)$ is used to
describe the images, and $(x',y,z')$ are the natural coordinates of
the OT (see Fig. \ref{OTandImaging}). The angle between $z$ (the
imaging beam axis) and $z'$ (the weak axis of the OT), or between
$x$ and $x'$ (one of the strong trapping axis of the OT), is
$\alpha$. We note $y$ the vertical axis, which is therefore the
other strong trapping axis of the OT.

From absorption images we obtain the Optical Depth, $OD(x,y)$, which
is linked to the atomic density $n(x,y,z)$ via:
\begin{equation}
OD(x,y)=\int\limits_{-\infty }^{+\infty }\sigma_{opt}\; n(x,y,z)dz
\label{defOD}
\end{equation}
Since the total number of atoms $N_{at}$ in the trap is proportional
to the total integral of $OD(x,y)$ as given in eq. (\ref{defOD}), we
deduce its value from:
\begin{equation}
N_{at}=\frac{1}{\sigma_{opt}}\int\limits_{-\infty }^{+\infty
}\int\limits_{-\infty }^{+\infty }OD(x,y)\;dxdy \label{Nat}
\end{equation}
To evaluate the integral in eq. (\ref{Nat}), we first perform a
numerical integration along $x$ of $OD(x,y)$. The function obtained
can be very well fitted by a Gaussian, and from the fit parameters
we deduce the value of the integral.

To obtain the peak density $n_{0}$ from $OD$ measurements, one needs
information on the atomic density distribution. The OT has a
cylindrical revolution symmetry around its weak axis, and the
density dependance on $y$ is decoupled from the ones on $x$ and $z$:
$n(x,y,z)=n(y)f(x,z)$. A cut of $OD(x,y)$ along $y$ gives therefore
the density distribution $n(y)=n(x')$. This distribution is well
fitted by a Gaussian, with a $1/e^{2}$ radius, $W_{t}$, equal to
$50$ ($36$) $\mu m$ for $f_{IR}$=20 (respectively 15) cm. In
addition, if the width of the density along the weak axis $z'$,
$\sigma_{z'}$, is much larger than $W_{t}$, we have in the vicinity
of the trap center:
\begin{equation}
n(x',y',z') \simeq n_{0}\;e^{\frac{-2(x'^{2}+y'^{2})}{W_{t}^{2}}}
\end{equation}
Using $x'=x \cos(\alpha)-z\sin(\alpha)$, we thus obtain:
\begin{equation}
n(x=0,y=0,z) = n_{0}\;e^{\frac{-2\sin(\alpha)^{2}z^{2}}{W_{t}^{2}}}
\label{n00z}
\end{equation}
By integrating eq. (\ref{n00z}) we can therefore link the maximal
value of the $OD$, $OD_{max}$, to the peak density:
\begin{equation}
n_{0}=\frac{OD_{\max }\sin (\alpha )}{\sqrt{\frac{\pi
}{2}}\sigma_{opt} W_{t}} \label{n0}
\end{equation}
This equation is valid as long as the imaging beam path through the
atomic cloud (along $z'$) is short compared to $\sigma_{z'}$. This
condition reads: $\frac{W_{t}}{\tan(\alpha)}<<\sigma_{z'}$.
Introducing the width of $OD(x,y)$ along the $x$ axis,
$\sigma_x=\sigma_{z'}\sin(\alpha)$, we finally get the condition
$\sigma_ x>>W_{t}$. Experimentally we measure $\sigma_x\simeq700$
$\mu m$, so that $n_{0}$ can be reliably deduced from eq.
(\ref{n0}).

\section{Experimental results}

The experiments described in this section were obtained with
$f_{IR}=15$ cm, except in subsection 3.3 where $f_{IR}=20$ cm.

\subsection{Accumulation of atoms in the OT}

Here we turn to our first main experimental result, $ie$ the
accumulation of cold metastable Cr atoms in the OT.

\begin{figure}
\resizebox{1\columnwidth}{!}{
\includegraphics{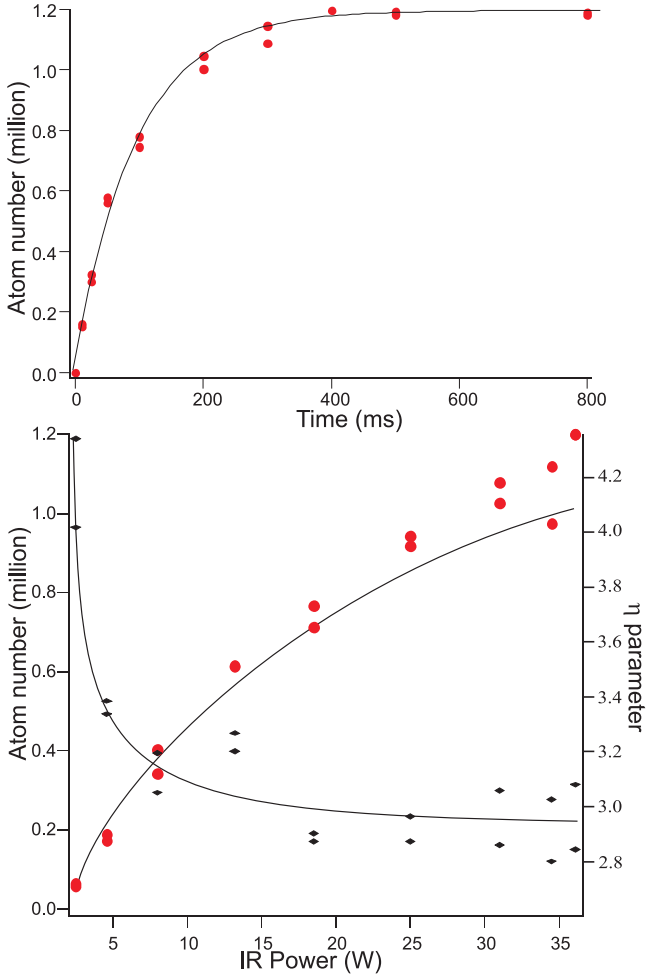}
} \caption{Top: Loading sequence into the OT at the optimal IR beam
focus position (see text), with the retroreflected beam, and at
maximum laser power. The line is the result of an exponential fit.
Bottom: dependance of the stationary number of atoms (circles) and
of the $\eta$ parameter (diamonds) - see text - with the trapping
laser power, which is set by AOM control. The two lines are guides
for the eye.} \label{OTloading}
\end{figure}

We show how fast the loading proceeds on the top part of Figure
\ref{OTloading}. To increase the number of trapped atoms, we have
retroreflected the IR laser beam, so that a total power of 70 W is
sent to the atoms. The ZS, the MOT, and the IR laser are switched on
for a given duration $t$. Due to eddy currents, it takes 20 ms for
the magnetic fields to die away after the power supply driving the
MOT coils is switched off. As a consequence, a 20 ms waiting time is
necessary before taking an image of the OT. During this waiting
time, the atom cloud has expanded along the $x$ axis. For each
loading time $t$, the total number of atoms is evaluated from eq.
(\ref{Nat}). A very short 1/e loading time of typically $\tau= 150$
ms is obtained.

From the image at $t=400$ ms, we deduce the following asymptotic
numbers: a total atom number $N_\infty=1.2$ $10^{6}$, and a peak
atomic density $n_0=1.2$ $10^{11}$ $cm^{-3}$. We want to emphasize
that the atomic density is much larger in the mixed trap (before the
20 ms expansion). Although is is difficult to extract the peak
density from the in situ absorption images due to the complicated
mixed trap geometry, we estimate that the longitudinal width is
about 10 times smaller before expansion, so that the in situ density
is about $10^{12}$ $cm^{-3}$.

In addition, we measured the temperature of the metastable atoms in
the OT, by analyzing the free expansion along the vertical $y$ axis
after the OT has been switched off. When the IR power is at maximum,
we obtain $T_{OT}=(120\pm10)$ $\mu K$. In the bottom part of Figure
\ref{OTloading}, we see that while the number of trapped atoms
increases with the trapping laser power, the parameter $\eta$, which
is the ratio between the OT depth (average theoretical value, see
eq. (\ref{Udip calcule})) and the cloud temperature, remains almost
constant ($\eta \simeq 3$) for large IR power.

\subsection{Loading dynamics}

To optimize the loading of the OT, we found that it is crucial that
the IR laser remain on during the whole MOT loading process: we
observe only little trapping if the dipole trap is turned on after
the MOT beams have been switched off. This indicates that the atoms
are directly injected in the OT from the MOT, rather than from the
MT. Indeed, an atom from the MOT traveling through the OT
experiences both the acceleration due to the OT potential and the
friction force of the MOT \cite{OT&MOT}. If friction is large
enough, and if it gets optically pumped to the metastable states
before it leaves the OT location, it will remain trapped in the OT.

\begin{figure}
\resizebox{1\columnwidth}{!}{
\includegraphics{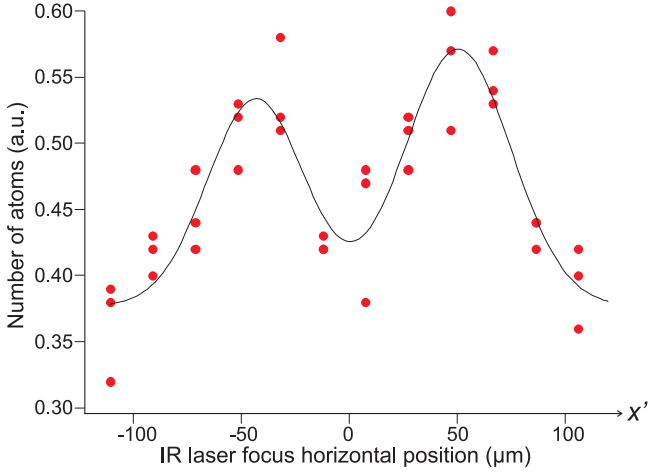}
} \caption{Dependency of the steady number of atoms $N{_\infty}$
with the IR laser focus position on the horizontal plane, with no
retroreflection. If the IR beam is crossing the MOT center ($x'=0$),
the atom number reaches a local minimum. The black line is a result
of a fit with two Gaussians. The two peak centers are separated by
100 $\mu$m.} \label{AlignmentH}
\end{figure}

To accurately discuss the loading physics and the alignment issue
between the OT and the MOT, we show on Figure \ref{AlignmentH} the
influence of the horizontal IR beam pointing on the stationary
number of atoms in the OT, $N{_\infty}$; there was no
retroreflection of the IR beam in this experiment to simplify the
analysis. We see that two optima are found for $N{_\infty}$. This
shows that the optimal situation is reached when the IR beam is
slightly off centered with respect to the MOT. Besides, we observed
that the loading time is longer when we set the IR beam at the
optimal positions ($x'=\pm50\mu m$, $\tau=240$ ms ), than if we set
it at the local minimum position ($x'=0$, $\tau_0=110$ ms). As the
loading time is set by the dominant loss rate, and as the inelastic
loss rate between metastable states should be higher for a larger
number of atoms, this rules out inelastic collisions between
metastable states as the limiting factor for the loading of the OT.
In addition, the loading time at the optimal position is equal to
the lifetime of the mixed trap at this position, which is measured
when the MOT beams are turned off after accumulation. This rules out
inelastic collisions with the MOT atoms as the limiting factor,
which is quite different from what was observed for ground state OT
loading from a MOT \cite{RbOTloading}. Finally, at low density, the
1/e lifetime (9s) of the pure OT is much larger than the loading
time, so that collisions with the background gas is not the limiting
factor either.

We therefore think that the dominant source of losses when the OT is
centered at B=0 is a consequence of Majorana losses: spin flips in
the vicinity of the zero B field position (which is as well the MOT
center) are responsible for the dominant loss process. Indeed the
high field seekers states are not trapped in the mixed trap, as they
are expelled along the weak trapping axis of the OT by the MT
gradient. When the IR beam is slightly off-centered, these losses
decrease, which accounts for the larger loading time that we
observe. At the same time the OT loading rate, proportional to the
superposition with the MOT, decreases too, which leads to the
experimental trade-off observed in Figure \ref{AlignmentH}.

For a matter of fact, the observed loss rate per atom when the OT is
centered at B=0, $1/\tau_0$, is in qualitative agreement with a
rough evaluation of the Majorana loss rate, $\Gamma_{maj}$. To
perform this evaluation, we calculate the probability per second for
an atom to cross the sphere of radius $d$ where a spin flip is
likely to happen, with $d=\sqrt{\frac{h\overline{v}}{\mu b_0}}$
\cite{MajoRate}. In this equality, $\overline{v}=(k_BT/m)^{1/2}$ is
the quadratic mean velocity along one axis in the mixed trap
($k_{B}$ is the Boltzmann constant, $m$ is the mass of the $^{52}$Cr
atom, $T$ is the temperature of the mixed trap); $\mu$ is the
magnetic moment of the $^5D_4$ state ($\mu=6\;\mu_B$), and $b_0$ is
the MT gradient. Along $z'$, the atom mostly feel the effect of the
MT, and a simple geometric argument gives:

\begin{equation}
\Gamma_{maj}=a\;\frac{2}{\tau_{MT}}\;\left(\frac{d}{W_t}\right)^2 \:
\label{Phi}
\end{equation}

where $\tau_{MT}$ is the typical MT period along $z'$, and $a$ is a
numerical factor of the order of one (for example, $a\simeq3$ for a
linear trap\cite{MajoRate}). Using the experimental value for $b_0$
(10 G.cm$^{-1}$), $W_t$ ($30$ $\mu$m), and $T\simeq100$ $\mu K$, we
obtain $d=4$ $\mu$m, $\tau_{MT}=30$ ms, and
$\Gamma_{maj}^{-1}=(800/a)$ ms, while $\tau_0=110$ ms.

\subsection{Parametric excitation spectra}

In this subsection, the IR laser is not retroreflected in order to
reduce systematic effects, since the exact size and shape of the
retroreflected beam is hard to measure. As noted above, $f_{IR}=20$
cm. In this situation, the measured temperature at full power is
70$\pm10$ $\mu$K, and the maximum number of atoms is 2 millions.

The measurement of an OT depth was performed by Friebel and
coworkers a few years ago \cite{QuadrupHansh} by modulating the trap
spring constant through the intensity of the trapping laser.
Depending on the modulation frequency $f_{m}$, the atom cloud gets
more or less heated, and atoms are driven out of the OT accordingly.
The parametric excitation spectra correspond to the number of atoms
remaining in the trap after a given modulation duration, as a
function of $f_{m}$. The characteristics of these spectra reflect
the trap potential shape.

We used the AOM controlling the trapping beam power to modulate the
OT depth. The Radio Frequency power sent to the AOM is set at 80\%
of its maximal value, so that an average power of 29 W is sent to
the atoms, and is modulated at a few kHz, with a peak to peak
amplitude of 20\%. To obtain the excitation spectrum of the
$^5$D$_4$ state, we first load the OT, and then modulate the IR
laser intensity for 40 ms. After repumping the atoms back to the
ground state, we measure $N_{at}$, as a function of $f_{m}$. An
excitation spectrum of the metastable state is shown in Figure
\ref{SimuExp}. We have also measured the excitation spectra of the
$^7$S$_3$ state, by first repumping the atoms, and then modulating
the OT depth.

\section{Simulation of the parametric excitation spectra}

\subsection{Classical 3D simulation of the parametric excitation process}

Since the trap potential (approximately Gaussian shaped) is not
harmonic, the parametric excitation spectra are both broadened and
shifted compared to the ones which would be obtained in a purely
harmonic trap, as noticed previously by other groups (see for
example \cite{TheseOHara}). In order to deduce the light shifts of
the levels of interest from the spectra described in subsection 3.3,
we performed a 3D classical Monte Carlo simulation, which also
allowed us to take into account the presence of the MT. The effect
of both elastic and inelastic collisions are not taken into account
in this analysis. We performed experimental parametric excitation
for various excitation duration, and found no change in the
experimental line shape, so that collisions can indeed be neglected.

The combined trap created by the MT and the OT is considered to be
the sum of two potentials. First a Gaussian 2D potential (in the
$x'y'$ plane) corresponding to the OT strong trapping axis, with a
depth $U_{0}$ ($U_{0}<0$) and a width $W$. Second, a linear 3D
potential, corresponding to the quadrupolar magnetic field created
by the MOT coils, described by the gradient $b_{0}$. During the
simulation, the trapping laser intensity is modulated for a duration
$\Delta t$, so that the atoms experience a time-dependent potential
$U(t)$  which reads as:

\begin{equation}
U(t)=U_{0}(1+f(t))e^{\frac{-2r^{2}}{W^{2}}}+\overline{m_J} g_J \mu_B
b_{0} \sqrt{x'^2+4y'^2+z'^2} \label{U(t)}
\end{equation}

where $r=\sqrt{x'^2+y'^2}$, $\overline{m_J}$ is the average magntic
quantum number ($\overline{m_J}\simeq2$, see \cite{Chi2}), and
$f(t)$ is the function describing the time evolution of the trap
depth.

In the Monte Carlo draw, the radial coordinate $r$, the longitudinal
one $z'$, and the velocity $V$ of the atoms, are selected according
to a Gaussian probability distribution, with respective 1/$e^2$
width $W{_t}$, $\sigma_{z'}$, and $2(k{_B}T/m)^{1/2}$ ($T$ is the
temperature of the cloud). We evaluate the final values of the
radial position ($r{_f}$), velocity ($V{_f}$), and the energy
$E{_f}=\frac{1}{2}mV{_f}^{2}+U_{0}e^{\frac{-2r{_f}^{2}}{W^{2}}}$. We
then evaluate the proportion of atoms remaining in the OT (which are
those with $E_{f}<0$) for different values of the modulation
frequency. We obtain the simulated parametric excitation spectra by
normalizing to the result obtained when no modulation is applied.

The values of $\Delta t$, $W{_t}$, $b_{0}$ and $W$ are the
experimental ones (40 ms, 50 $\mu$m, 10 G.cm$^{-1}$, 55 $\mu$m).
$\sigma_{z'}$ is evaluated from images of the mixed trap (
$\sigma_{z'}\simeq280$ $\mu$m). The function $f(t)$ reproduces the
experimental laser intensity modulation. The simulated spectra are
quite insensitive to the temperature $T$. In the simulation of
Figure \ref{SimuExp}, $T=120$ $\mu$K.

\subsection{Numerical results}
We computed spectra for various value of $U_{0}$, using $U_{0}$ as
the only adjustable parameter. The simulated spectra are always
narrower than the experimental ones, and we focused on the position
of the central excitation frequency: we look for the value of
$U_{0}$ that gives a spectrum centered at the experimental
resonance.

\begin{figure}
\resizebox{1\columnwidth}{!}{
\includegraphics{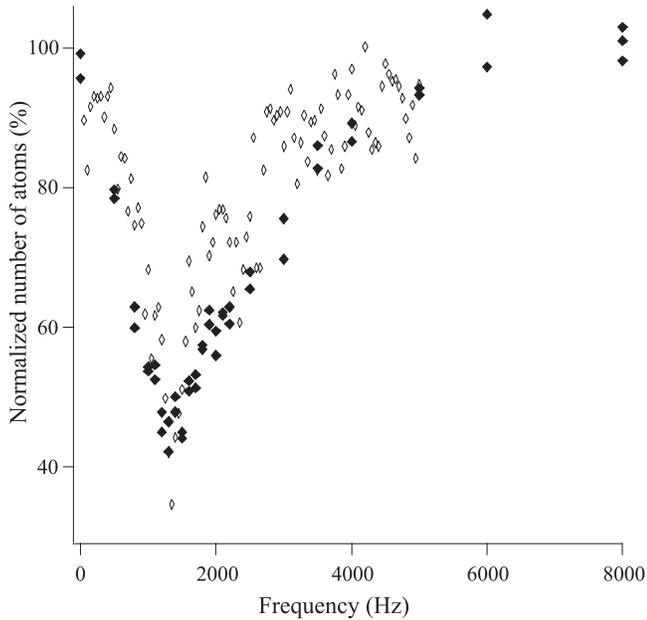}
} \caption{Experimental and best adjusted simulated excitation
spectra (open diamonds) for the $^5$D$_4$ state. The normalized
remaining number of atoms is plotted versus the trap modulation
frequency.} \label{SimuExp}
\end{figure}

The experimental and the best central frequency adjusted theoretical
spectra are shown for the $^5$D$_4$ state in Figure \ref{SimuExp}.
From this best adjustment, we deduce the following OT depth in MHz:
$U_{0\, ^{5}D_{4}}=-(3.2\pm0.6\pm0.6)$ MHz. The first uncertainty of
20\% is our estimate of the fit uncertainty. The second is a result
of the 10\% accuracy on the waist measurement. We have applied the
same procedure to spectra obtained for atoms in the ground state
(for which $g_J=2$), and obtained $U_{0\,
^{7}S_{3}}=-(3.8\pm0.7\pm0.7)$ MHz. As theoretical evaluations have
already proved to be in good agreement with experimental values for
the ground state light shifts \cite{TheseGriesmaier}, the comparison
of $U_{0\, ^{7}S_{3}}$ with calculations in the next section will
allow us to test the validity of our procedure.

\section{Theoretical evaluations of the light shifts and comparison with the experiment}
\label{sec:5}

We now turn to our theoretical estimates for the light shifts to
compare with the experimental data.

When a given atomic level $i$ having an energy $\hbar\omega _{i}$ is
coupled by a laser through dipolar electric transitions to other
states $n$ having energies $\hbar\omega _{n}$, its energy is shifted
due to the presence of the AC laser electric field. This light shift
depends on $I_{L}(r)$, the local laser intensity, on $\omega
_{ni}=\omega _{n}-\omega _{i}$ and on $\gamma _{ni}$ (respectively
the angular frequency and the cycling rate of the transition between
level $i$ and level $n$), and on $\omega $ the laser angular
frequency. In fact, the different Zeeman sublevels- corresponding to
the magnetic quantum number $M_{Ji}$ - are coupled differently to
the excited states, and the light shift not only depends on the
total angular momentum $J _{i}$ but as well on $M_{Ji}$, and on the
laser polarization. To simplify the discussion we take a nuclear
spin equal to zero, which is the case for the bosonic $^{52}$Cr
atom. The coupling rate between the sublevel ($i$, $J_{i}$,
$M_{Ji}$) and the excited state ($n$, $J_{n}$, $M_{Jn}$) is given by
a 3J coefficient, and one must distinguish between a coupling
towards more energetic states ($\omega _{ni}>0$) or toward less
energetic states ($\omega _{ni}<0$):
\begin{eqnarray}
\gamma _{M_{Ji},n} = \gamma _{ni}\left( 2J_{n}+1\right) \times
C_{J_{i},M_{Ji},J_{n},sign(\omega _{ni})\times q}^{2}
\end{eqnarray}

In these formula, $q$ gives the laser polarization ($q=-1,0,+1$),
and the 3J coefficient $C_{J_{i},M_{Ji},J_{n},q}$ is equal to
$\left(
\begin{array}{ccc}
J_{i} & 1 & J_{n} \\
-M_{Ji} & -q & M_{Ji}+q
\end{array}\right)$.

The light shift can therefore be written to the lowest order in
perturbation theory as \cite{GrimmReview}:
\begin{equation}
U_{i,M_{Ji}}(r)=-3\pi c^{2}I_{L}(r)\sum\limits_{n}\frac{sign(\omega _{ni})\gamma _{M_{Ji},n}}{%
\omega _{ni}^{2}(\omega _{ni}^{2}-\omega ^{2})}  \label{Udip
general}
\end{equation}

Some simplification occurs if the state $i$ and all the states it is
coupled to belong to true spectroscopic terms, which means that they
are fully characterized by their total orbital momentum, $L$, and
spin, $S$, in addition to their total momentum $J$. First, the
cycling rate $\gamma _{ni}$ does not depend on the fine structure.
Then, if the laser frequency is red detuned enough so that all the
detunings are large compared to the fine structure splitting, we can
use the following sum rule over the 3J coefficients for further
simplification:
\begin{equation}
\sum\limits_{J_{n}=J_{i-1},J_i,J_{i+1}}(2\times
J_{n}+1)C_{J_{i},M_{Ji},J_{n},q}^{2}=1,\forall q
\end{equation}
Finally, if the state $i$ is only coupled to more energetic states,
the light shifts depend neither on $M_{Ji}$, nor on the laser
polarization. This explains why the fine structure can be ignored
when evaluating the light shift of the ground state in various
instances.

However, this simplification is not valid anymore if either the
state $i$ or the levels it is coupled to, are not pure spectroscopic
terms (which is the case when $\gamma _{ni}$ depend on $J_{n}$).
This is indeed the case for Chromium. The coupling of the ground
state $^7S_3$ to excited $P$ states depends on the fine structure of
these excited states. The effect is even larger for the $^5D$
metastable states, which are themselves not pure spectroscopic terms
(as clearly shown by their non zero coupling to the $^7P$ state).

To evaluate the light shifts of the levels of interest at the center
of the IR laser beam from eq. (\ref{Udip general}), we used the
spectroscopic data available from NIST \cite{NIST} for the atomic
level energies and the coupling rates, and we took for $I_{L}(r)$
the experimental laser intensity peak given by the measurements of
the laser power (P=29 W) and waist (W=55 $\mu$m): $I_{L}=2\times
P/(\pi W^{2})$. Because the parametric excitation is performed in
the mixed trap where the magnetic field direction is not constant,
there is no good choice for $q$. As the laser polarization is
linear, the polarizations seen by the atoms in the mixed trap change
between a $\pi$ polarization ($q=0$), and an equal mix
-$\sigma_{inc}$ polarization- of $\sigma$+ ($q=1$) and $\sigma$-
($q=-1$) polarizations, corresponding to a local B field
respectively orthogonal and parallel to the laser propagation axis.
We give below (in MHz) the results obtained for both states:
\begin{eqnarray}
U_{0 \, ^{7}S_{3},M} & =_{\sigma_{inc}}& -2.525-0.004\times M^{2}  \nonumber \\
& =_{\pi} & -2.574+0.008\times M^{2}\nonumber \\
U_{0 \,^{5}D_{4},M} & =_{\sigma_{inc}} & -1.925-0.015\times M^{2}\nonumber \\
& = _{\pi} & -2.220+0.03\times M^{2}\label{Udip calcule}
\end{eqnarray}
For both polarizations, the quadratic dependance on $M$ is a result
of properties of the 3J coefficients \cite{SumRule}.

As discussed above, there are small differences for the light shifts
of the ground state Zeeman sublevels, with a maximum of the order of
3\%. We want to stress that this energy splitting has to be taken
into account when evaluating spinor BEC phase diagrams
\cite{SpinorBEC} in an OT. Nevertheless, the fact that these energy
shifts remain small in percentage allows us to compare the
theoretical results and the experimental one.

The situation is quite different for the $^5D_4$ level: up to 21\%
differences are expected between the light shifts of the $M_{J}=4$
and $M_{J}=0$ sublevels for a $\pi$ polarization. Therefore, an
accurate comparison between theory and experience would require to
work with a polarized atomic sample, but we cannot achieve this
polarization experimentally, as any experimentally accessible
optical excitation from the $D$ metastable states drives the atoms
to other levels. Besides, it should be noted that spectroscopic data
concerning the couplings of the metastable state are less complete
than the ones of the ground state.

We see that the theoretical values of eq. (\ref{Udip calcule}) are
below our experimental light shifts measurements for both the
$^7S_3$ and the $^5D_4$ state, although within the error bars. We
are still investigating this issue. One possibility is that the IR
laser mode at the atoms location is affected by systematic thermal
effects induced by the crossing of the viewports of the experimental
chamber.

In conclusion, we have optically trapped metastable Cr atoms, using
a new accumulation procedure. We have identified the leading loss
mechanism which limits the accumulation of atoms. The trap
characterization shows that we have obtained a promising starting
point to reach Cr quantum degeneracy following a new strategy. The
measurement of the trap depth is in reasonable agreement with
calculations based on available spectroscopic data. For atoms having
a complicated electronic structure such as Cr, our calculations
stress the absence of energy degeneracy among the different
subZeeman states in an OT, even if it is created by a far red
detuned laser.


\begin{thebibliography}{}
\bibitem{SumRule}
$C_{J_{i},M_{Ji},J_{n},0}^{2}$ , or the sum
($C_{J_{i},M_{Ji},J_{n},1}^{2}+C_{J_{i},M_{Ji},J_{n},-1}^{2}$), do
not include any term proportionnal to $M_{Ji}$, for
$J_n=J_{i-1},J_i,J_{i+1}$.
\end{thebibliography}
\end{document}